# Chapter 12: Hyperfine-mediated transport in a one-dimensional channel

M. H. Fauzi[1] and Y. Hirayama[2,3,4]

[1]Research Center for Physics, Indonesian Institute of Science, South Tangerang City, Banten 15314, Indonesia

[2]Center for Spintronics Research Network, Tohoku University, Sendai 980-8577, Japan

[3]Department of Physics, Tohoku University, Sendai 980-8578, Japan

[4]Center for Science and Innovation in Spintronics, Tohoku University, Sendai 980-8577, Japan

**Abstract.** We survey some recent accumulated body of works on hyperfine-mediated transport in a confined one-dimensional channel, realized typically by electrostatic gating. Our review begins with how the spin-polarized edge current can be used as a means to dynamically generate and detect an ensemble of nuclear spins in the channel. We show how the transmission of spin-polarized edge in the presence of nuclear polarization could provide a convenient way to explain various NMR lineshapes observed in experiments. We discuss recent attempt to electrically detecting NMR at moderate to low magnetic fields that would particularly be helpful to address some fundamental physics problems in quantum transport such as the microscopic nature of fractional conductance developed below the last integer plateau. For a more applied-oriented, the developed NMR can be turned to probe an ultra-low level strain modulation in the channel owing to differential thermal contraction between metal gates and a semiconductor via nuclear quadrupole interaction.

## 1. Introduction

Nuclear magnetic resonance (NMR) is a spectroscopic tool which probes structural and dynamical properties in materials non-invasively. Developed in 1940s during the second World War, the technique has found its application in a broad range of disciplines of science, engineering, and medicine. NMR typically operates at radio frequency spectrum to perturb nuclear spin magnetization in materials.

Unlike electron spins, nuclear spins are paramagnetic with a Curie temperature lies in μK regime. This weak nuclear spin magnetization combined with low efficiency of conventional detection technique makes NMR signal inherently insensitive. Conventional NMR detection relies on Faraday induction by a pick-up resonator placed nearby the probed material. A carefully design resonator with a high $Q$-factor is necessary to have a good detection sensitivity. Despite the improvement, the conventional NMR is still limited to dimensions with at least a micro-meter size in volume, limiting the applicability to nanostructures[Lee01].

| Technique | Detection method | Min. Sensitivity |
|---|---|---|
| Conventional NMR | Faraday induction | $10^{12}$ spins [Lee01] |
| Resistively-detected NMR | Electrons | $10^{6}$ spins [Ono14] |
| Optically-detected NMR | Photoluminesence | $10^{6}$ spins [Chekhovich13] |
| Force microscopy | Ferromagnetic cantilever | $10^{4}$ spins [Rugar04] |

Table 12.1. List of several existing approaches to detecting nuclear magnetic resonance.

Several alternative approaches have been developed throughout the years to detect a smaller number of nuclear spins as listed in Table 12.1. Unlike the conventional detection, the detection sensitivity of those newly developed techniques do not scale with material dimensionality and thus applicable to nanostructures. Of particular interest is resistively-detected NMR discovered in the late 80s by von Klitzing group on a 2D quantum Hall system [Dobers1988]. The topic has been a subject of several review papers, mostly emphasized on the hyperfine-mediated transport in a 2D quantum Hall system [Li2008, Gervais2009, Hirayama2009]. Our current review here is focused on similar topic observed in a quasi 1D semiconductor system including its recent development. The quasi-1D system typically realized in a quantum point contact is an attractive platform to study a number of fundamental physics and serves as a basic building block for future quantum nanoelectronics [Duprez2019].

## 2. Quantum point contacts

A quantum point contact (QPC) is a tunable quasi-1D electron waveguide, analogous to optical waveguide [vanWees1988, Wharam1988]. The point contact is typically realized by imprinting an electrostatic potential on a two-dimensional electron gas (2DEG) through external gates hovering above it. The number of conductance modes passing through the QPC can be precisely controlled by applying negative bias voltages to a pair of split metal gate patterned on the surface [Thornton1986]. Additional gate or more complex architectures could be incorporated to modify confinement potential profile for instance as schematically displayed in the inset of Fig. 12.1.(a). Although other techniques exist such as trenched QPCs, a gate-defined QPC embedded in a 2DEG is still a preferable choice since it gives a smoother boundary, ensuring tunable ballistic transport over the channel.

Such experimental realization is demonstrated in Fig. 12.1(a) where we plot a semi-log of a linear zero-field conductance $dI/dV$ in the unit of $2e^2/h$ as a function of $V_{SG}$. The center gate is kept at $V_{CG} = 0$ V throughout the measurement. Initially the conductance is very high, but then with increasing a negative bias voltage to the split gate pair, we see a sharp drop in the conductance, marking the transiton from 2D to 1D transport. A number of quantized conductance plateau, in a step of $2e^2/h$, is seen.

The conductance *G* through a point contact is simply given by the sum of transmission probability *T(E)* from all occupied channels *n* below the Fermi level [Buttiker1990]

$$G = \frac{2e^2}{h} \sum_n \int T_n(E) dE$$

Intuitively speaking, when the split gate voltage is swept, the Fermi level continously moving through each energy level until the point contact gets pinched off. The current allows to drop when it coincides with the energy level but ceases to drop when it lies in between the two adjacent energy levels and hence the conductance gets quantized. The plateau length thus gives us a sense of how large the subband spacing is between the two adjacent energy levels.

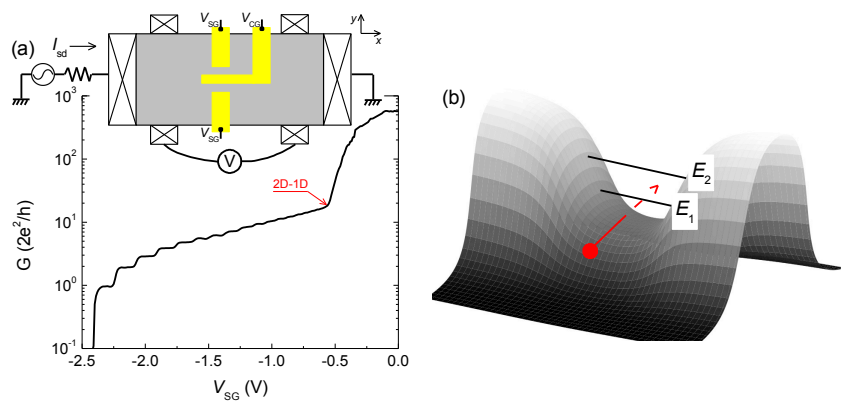

Fig. 12.1. (a) A semi-log plot of zero-field conductance trace as a function of split gate bias voltage. A quantum point contact is defined by applying a negative bias to a pair of split gates. Additional center gate in between the pair is used to modify the potential confinement profile. A transition from 2D to 1D transport is indicated by the red arrow. Inset shows device schematic (not to scale) and measurement setup. Adapted with permission from [Fauzi2018]. (b) Illustration of a saddle point potential with a quantized energy level formed in the point contact.

### 3. 1D Magnetotransport

Now what happens when we plunge the device in a perpendicular magnetic field?. The 1D energy spectrum would definitely be modified with several additional terms appear in the equation. For a non-interacting with the subband index $n$ and electron spin $S_z$, the energy spectrum at $k = 0$ is given by

$$E(n, S_z) = \left(n + \frac{1}{2}\right)\hbar\omega + g_n^*\mu_B B S_z + A I_z S_z$$

here the first term describes the magneto-electric subband mixing with $\omega = \sqrt{\omega_y^2 + \omega_c^2}$ . $\omega_y$ and $\omega_c = eB/m^*$ are the lateral confinement potential and cyclotron frequency, respectively. The second term is the electronic Zeeman energy with an effective Lande $g_n^*$-factor whose magnitude is subband-dependent. The last term is the contact hyperfine interaction between a nuclear spin and an electron. Due to technological maturity in fabrication processes, GaAs is still a primary testbed for quantum nanoelectronics. GaAs, apart from electron spin degree of freedom, has another important spin degree of freedom namely nuclear spin that would influence the transport as well.

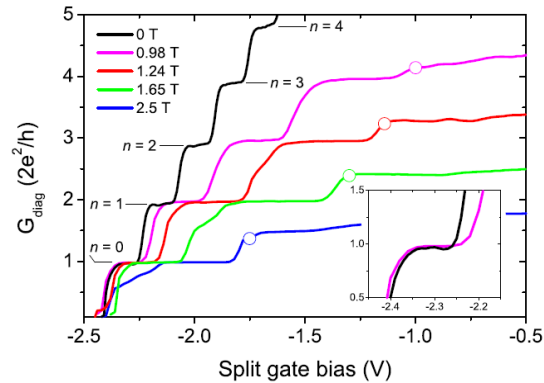

Fig. 12. 2. Diagonal conductance as a function of split gate bias voltage measured at several perpendicular magnetic field. The inset shows comparison of the first integer plateau measured at zero and 0.98 T field. Reprinted with permission from [Fauzi2018]

The semiconductor hosts three active nuclei whose properties are listed in Table 12.2. Owing to a relatively high hyperfine constant, thanks to its *s*-wave like conduction band structure, the maximum Overhauser field experienced by an electron spin could be as high as -5.3 Tesla [Paget1997], provided that all nuclear spins were polarized. As we have discussed earlier that the nuclear spins are paramagnetic, so that its polarization remains small even at cryogenic temperature and plunge them in high magnetic field. For example at $T$ = 100 mK and $B$ = 10 Tesla, the thermal polarization of $^{75}$As nuclei is only a mere 3.5%. However, we can increase the polarization by the so-called dynamic nuclear polarization, taking advantage of angular momentum exchange between electron and nuclear spin within the point contact. In fact, when the condition is optimized, Dixon et al estimated an esemble of nuclear spin polarization of as high as 85% can be generated dynamically in a GaAs based quantum point contact [Dixon1997]. Such high polarization is possible since the electron dipole moment is three order of magnitude higher than its counterpart.

| Isotopes | $^{69}$Ga | $^{71}$Ga | $^{75}$As |
|---|---|---|---|
| Abundance (%) | 60.1 | 39.9 | 100 |
| Spin number, $I_z$ | 3/2 | 3/2 | 3/2 |
| Gyromagnetic ratio, $\gamma_{\scriptscriptstyle n}$ (MHz/T) | 10.428 | 13.021 | 7.29 |
| Hyperfine constant, A ($\mu$eV) | 38 | 49 | 46 |
| Max. Overhausef field, $B_N$ (T) | -1.37 | -1.17 | -2.76 |
| Quadrupole moment, $Q$ (x $10^{-25}$ cm$^2$) | 1.99 | 1.07 | 2.7 |

Table 12.2. Properties of stable nuclear isotopes in GaAs semiconductor.

The effect of magneto-electric subband mixing on the conductance can be readily seen in Fig. 12.2. First, the number of plateau decreases with increasing the field but the plateau length gets extended. Second, comparing the conductance traces for each selected field, one can immediately see that the length of each conductance plateau, integer or half-integer plateaus, reduces with lowering the subband index *n*. This can be understood since as the channel width gets narrower, the confinement potential $\omega_y$ component gets bigger with respect to $\omega_c$ component. In this case, the plateau width is mostly determined by the electrostatic confinement potential $\omega_y$.

When the cyclorton energy is much larger than the confinemet potential ($\omega_c \gg \omega_y$), the transport enters quantum Hall effect regime. Quantum Hall provides a natural realization of chiral conducting edge state protected by the gapped bulk state (incompressible). In fact, the chiral current running along the edge of two-dimensional plane is considered to be an ideal representation of 1D conductor [Duprez2019]. The number of edge states (2$v$) can be counted from the quantized Hall conductance ($G = 2ve^2$/h) as shown in Fig. 12. 3(a). Several even bulk filling factors $v_b$ are indicated in the panel. For example, for filling factor $v_b = 4$, there are 4 edge states and each is separated and protected by incompresible state. The filling factor itself is controlled by the electron density *n* and magnetic field *B* which satisfies $v = eB/nh$. An excellent review on the edge states and its historical context is given by R. Haug [Haug1993].

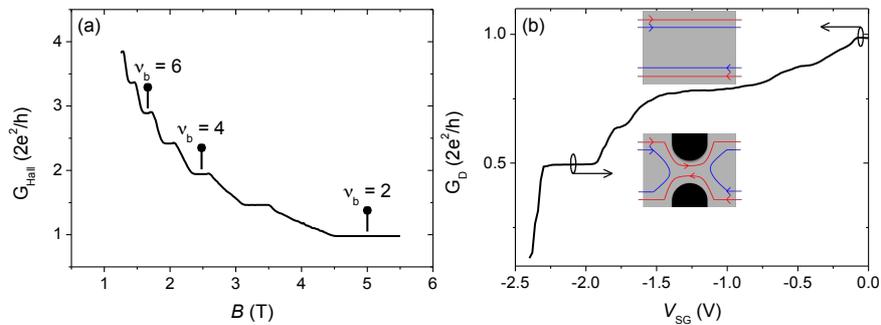

Fig. 12.3. (a) Hall conductance trace as a function of magnetic field. Even bulk filling factors are indicated. (b) Diagonal conductance trace as a function of split gate bias voltage measured at *B* = 5 Tesla and *T* = 100 mK. Inset shows the corresponding schematic of edge states at each respective electronic state. Red and blue lines in the inset indicates the up-spin and down-spin electron, respectively. Reprinted with permission from [Fauzi2018]

Once the bulk filling factor is determined, we can squeeze the point contact and control the number of edge state transmitted through. This is exactly accomplished in Fig. 12.3(b). Here the diagonal conductance is recorded as a function of split gate bias voltage $V_{SG}$ at $B$ = 5 Tesla ($v_b$ = 2) and $T$ = 100 mK. The bulk 2DEG is set at filling factor $v_b$ = 2 where the up- and down-spin edge channel are available for transmission (see the inset). Applying a negative voltage to $V_{SG}$ then allows us to selectively transmit the up-spin edge channel and reflect the down-spin edge channel. As we shall see in the next section, this particular control is necessary to promote inter edge scattering leading hyperfine-meditated transport.

### 4. Dynamic nuclear polarization in GaAs point contacts

At the heart of resistively detected NMR (RDNMR) is a contact hyperfine interaction between electron and nuclear spins as discussed earlier. The interaction is reciprocal and its Hamiltonian can be expressed as

$$H = A\mathbf{I} \cdot \mathbf{S} = \frac{A}{2}(I_-S_+ + I_+S_-) + AI_zS_z$$

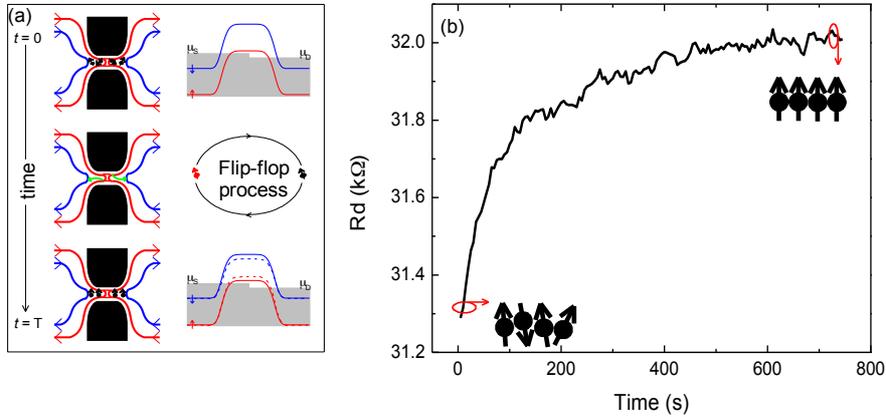

Fig. 12.5. (a) Schematic illustration of hyperfine-mediated inter edge scattering events in the point contact, leading dynamic nuclear polarization build up. (b) Time trace of diagonal resistance during dynamic nuclear polarization illustrated in panel (a). Adapted with permission from [Fauzi2017]

depending on the band structure, the hyperfine constant $A$ can be either isotropic or anisotropic. A system with predominantly s-wave band structure like $n$-type GaAs, its hyperfine constant is isotropic. The notation $\boldsymbol{I}$ and $\boldsymbol{S}$ denote the usual nuclear and electron spin operator, respectively. The Hamiltonian is usually broken down into two terms. The first term, the spin flip-flop term, describes dynamical process by allowing them to exchange their angular momentum. The process provides a pathway for dynamic nuclear polarization build-up as well as relaxation toward equilibrium. The second term is a static term, out of which each party would experience additional magnetic field. In fact, this last term is important since it makes the electrical detection of nuclear spin possible.

The simplest example to induce dynamic nuclear polarization in the point contact is by setting the bulk filling factor to $v_b = 2$ and the point contact to $v_{QPC} < 1$. This way, inter-edge state scattering involving two opposing electron spin would take place as illustrated in Fig. 12.5(a). The electron spin reversal from the outer edge channel with up-spin to the inner edge channel with down-spin is mediated by the hyperfine interaction to conserve its angular momentum. As a result, nuclear spin polarization parallel to the field direction is generated dynamically within the point contact.

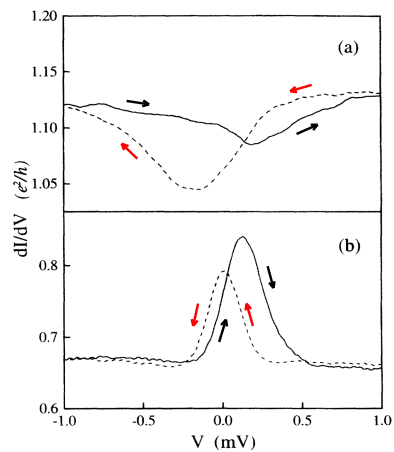

Fig. 12.6 Conductance traces through a single quantum point contact as a function of dc source-drain voltage with two opposing sweep directions for (a) $v_{QPC} > 1$ and (b) $v_{QPC} < 1$. Reprinted with permission from [Wald1994]

The dynamic polarization is manifested in the increase of the diagonal resistance (or conductance decrease) as time progresess as depicted in Fig. 12.5(b). An quasi-ac source-drain current is kept constant at $I_{sd}$ = 10 nA throughout the measurement. The resistance is increased exponentially with a characteristic rise time of about 150 seconds and in principle is current density dependence. This slow dynamic is a typical characteristic of hyperfine mediated transport.

An obvious question would be why the resistance increases rather than decreases with time?. To answer the question we need to understand how the point contact potential barrier height seen by electrons is modified by the presence of nuclear spin polarization or the Overhauser field $B_N$ to be exact. The potential barrier $U$ seen by electrons consists of two terms as follow

$$U(S_z) = U_0 + g^* \mu_B B^* S_z$$

here $U_0$ is the bare potential barrier, which is geometric-dependent. The second term is the modified Zeeman term with $B^* = B + B_N$ stands for the effective magnetic field felt by the up- or down-electron spin. The Overhauser field created by the upward nuclear polarization acting back on the electron spin is negative and the electronic Zeeman energy is reduced. Consequently, when $v_{QPC} < 1$ the up electron spin would see an increase in the potential barrier and thereby its transmission probability across would lower. This is the reason why we see an increase in the resistance during current-induced dynamic nuclear polarization in Fig. 12.5(b). For $v_{QPC} > 1$, the response would be the opposite namely the down spin electron would see the potential barrier gets lowered.

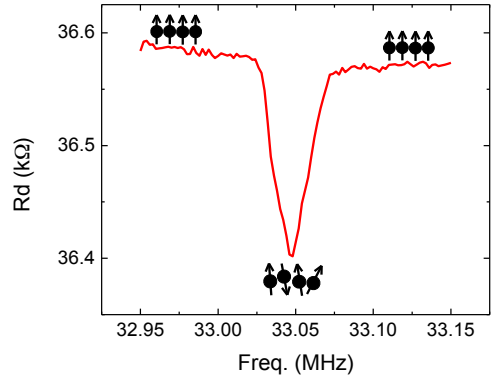

Fig. 12.6. $^{75}$As RDNMR of a point contact as a function of rf frequency. The spectrum is swept with increasing the frequency at a rate of 100 Hz/s and amplitude of -30 dBm. The spectrum is measured at $B$ = 4.5 T and $T$ = 300 mK. Reprinted with permission from [Fauzi2017]

An early study by Kane *et al* in 1990s used a spin-diode method instead to prove hyperfine-mediated transport [Kane1992]. They observe hysteretic feature in the I-V curve and attributed the effect as a signature of dynamic nuclear polarization. The idea was then further extended to a double and single quantum point contact by Wald *et al* from which similar feature was obtained as depicted in Fig. 12. 6 [Wald1994].

Although all experimental demonstrations in the point contact have undoubtly pointed out to the hyperfine-mediated transport, however its theoretical studies are still lagging behind. Only recently Anirban *et al* [Singha2017] has successfully reproduced the main feature observed in Fig. 12.6. by casting the hyperfine interaction directly into Landauer-Buttiker equation. Similar footstep is used as well by Stano *et al* but in a parallel magnetic setting [Stano2018].

## 5. RDNMR lineshapes in a quantum point contact

Although we have discussed some of the signature of hyperfine-mediated transport in the previous section, however a definite smoking-gun evidence must be coming from NMR spectrum. In a typical experiment, dynamic nuclear polarization is immediately followed by sweeping radio frequency magnetic field across Larmor frequency of a nuclei, in this particular case to Ga or As nuclei. The rf is applied to a home-made 0.3 mm thick copper coil wounded around the device multiple times. The rf sweep itself operates in a continuous wave mode where the rf irradiation time far exceeds nulear spin coherence time on the order of milisecond. Note that coherent oscillation of nuclear spin can be measured using various rf pulse protocols as demonstrated in a number of reports [Yusa2005, Ota2007, Corcoles2011].

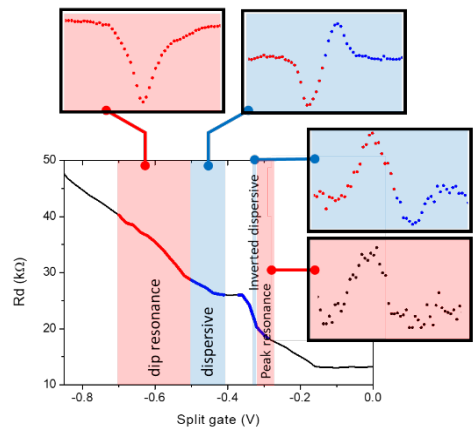

Fig. 12.7. Various emerging RDNMR lineshapes observed in the point contact in the so-called strong tunneling regime. The spectrum is measured at *B* = 4.5 T and *T* = 300 mK. Adapted with permission from [Fauzi2017]

A typical RDNMR signal of a point contact is displayed in Fig. 12.6. The generated in-plane rf magnetic field couples to the nuclear spin in such a way that when the rf hits the Larmor frequency, the resistance drops of about 200 Ω since the nuclear polarization is destroyed. The nuclear spin gets repolarized when the rf passes the resonance point. One thing worth mentioning that stands out from and in contrast to RDNMR in 2D systems is that it responses well even to a very low rf power of -30 dBm or less. Similar responsiveness is also observed in Nuclear Electric Resonance of a quantum point contact [Miyamoto2016].

Since the hyperfine field directly influences the transmission probability of a given spin-edge channel through the point contact as we explain in the previous section, one can easily understand how various RDNMR lineshapes emerged as well as their associated spin-flip scattering events in the point contact. This is surprisingly in contrast to the 2D systems, where the lineshapes are still poorly understood to date.

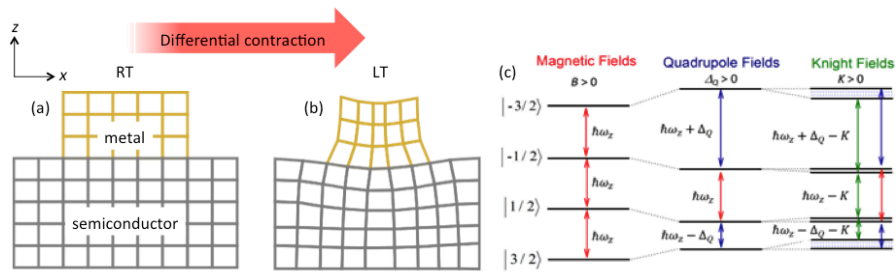

Fig. 12.8. (a)-(b) Differential thermal contraction between metal and semiconductor when cooled from room (RT) to low (LT) temperature. (c) Energy diagram of $I = 3/2$ nuclear spin system in the presence of external magnetic field (left), non-zero quadrupole splitting (middle), and Knight fields (right). Reprinted with permission from [Noorhidayati2020]

Motivated by the lack of understanding of the lineshapes, Fauzi *et al* study various RDNMR lineshapes in the point contact and the findings are summarized in Fig. 12.7. The measurements are carried out in a strong tunneling regime and observe four different lineshapes. In the vicinity of vqpc = 1 plateau, a dispersive (inverted dispersive) RDNMR lineshape is emerged, resembling those found in the 2D system [Desrat2002]. Outside this regime, the resonance turns into an usual dip or peak structure. The peculiar dispersive lineshapes emerge due to simultanous occurence of two spin-flip scattering events, giving rise to two localized regions with opposite nuclear spin polarization. Although both of them are in contact with electrons in the point contact, but they polarize in a region with different degree of electron spin polarization.

## 6. Structural lattice deformation

Another important aspect of a high-spin nucleus ($I > 1/2$) is that it has a quadrupole moment that can sensitively sense electric field gradient (EFG) from the surrounding environment. One major source of EFG is coming from structural deformation that forcibly disturbs charge arragements in the host crystal. Anisotropic strain variation of less than $10^{-4}$ can be detected by a quadrupole nuclei [Sundfors1969].

The electric field gradient $V$ (EFG) and strain $\varepsilon$ are linked by elastic tensor matrix $S$ as follow.

$$\begin{pmatrix} V_{xx} \\ V_{yy} \\ V_{zz} \end{pmatrix} = \begin{pmatrix} S_{11} & S_{12} & S_{12} \\ S_{12} & S_{11} & S_{12} \\ S_{12} & S_{12} & S_{11} \end{pmatrix} \begin{pmatrix} \varepsilon_{xx} \\ \varepsilon_{yy} \\ \varepsilon_{zz} \end{pmatrix}.$$  1

Due to GaAs crystal symmetry, $2S_{12} = -S_{11}$ [Taylor1959]. First-order quadrupole splitting ($\Delta_Q \ll E_Z$), with the magnetic field oriented parallel to the principal $z$-axis, is given by

$$\Delta_Q = \frac{eQS_{11}}{2h} V_{zz} = \frac{eQS_{11}}{2h} \left[ \varepsilon_{zz} - \frac{1}{2} \left( \varepsilon_{xx} + \varepsilon_{yy} \right) \right],$$  2

here the new notation $S_{11}$ is the elastic tensor component. The first order quadrupole splitting modifies the 3/2-spin nuclear energy level into three non-equidistance levels as depicted in Fig. 12.8(b), namely central and two sattelite transitions. It is important to note that for the first oder perturbation, the central transition is not affected by the quadrupole splitting but the Knight field. We will come back again to this important point and put to use in section 9.

A common example of structural deformation involves nano-meter metal gate patterning on GaAs semiconductor surface used to restrict electron/hole movement into reduced dimension. Since the metal and semiconductor have different coefficient of thermal expansion, naturally strain would develop at the interface when the device is cooled to cryogenic temperature as schematically displayed in Fig. 12.8(a). The strain then propagates down to the active 2DEG layer embedded in the semiconductor and may alter its electrical and optical properties. However, the imprinted strain distribution in the nano-gate/semiconductor plat-form is largely unknown, partly due to the fact that the initial strain from differing thermal expansion coefficients has to be experimentally determined.

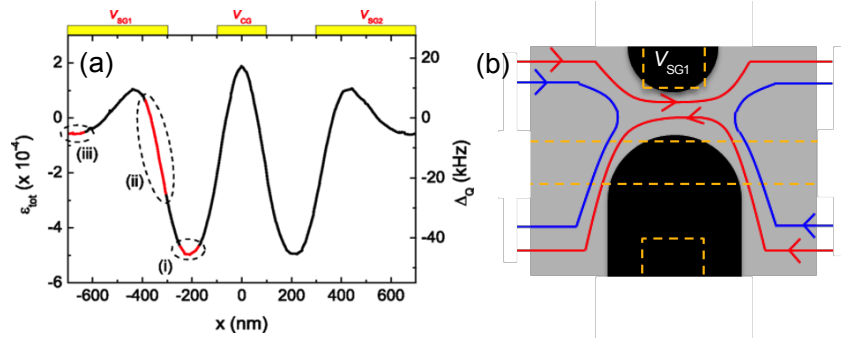

Fig. 12.9. (a) Simulated lateral strain profile using COMSOL mutliphysics and the corresponding quadrupole splitting at the center of quantum point contact located 175 nm below the surface. Reprinted with permission from [Fauzi2019]. (b) Race track-like schematic of spin edge current. The yellow dashed line indicates the gate layout hovering above the 2DEG. The black region indicates the depleted region.

For a practical purpose, let consider the case for a triple-gated quantum point contact device shown in Fig. 12.1(a). An elastic model calculation allows us to evaluate the strain pattern imprinted 175-nm below the surface where the channel is located as depicted in Fig. 12.9(a). The initial strain developed at the interface is taken to be 5 x 10⁻³ to match the experimental data shown in Fig. 12.10. Since the strain profile is mirror symmetric, one can target only half the section ($x < 0$) and deplete the other half. We identify three representative regions to focus on, (i) – (iii), as indicated in Fig. 12.9(a). To guide the edge channel to pass through the selected region, we can use gate bias tuning technique as schematically displayed in Fig. 12.9(b).

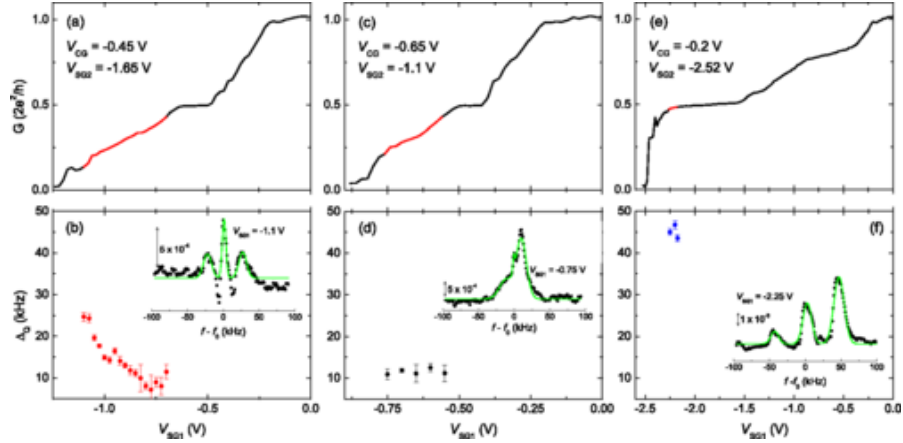

Fig. 12.10. (top panels) magneto-conductance traces and the corresponding quadrupole splitting along the conductance marked by the red line. The spectra are measured at $B$ = 4.5 T and $T$ = 300 mK. Inset in each bottom panel shows a represented RDNMR spectrum. Reprinted with permission from [Fauzi2019]

Since this is kind of a blind experiment, many gate bias tuning trials are needed. Region (ii) can be accessed by tuning $V_{CG}$ = -0.45 V and $V_{SG2}$ = -1.65 V fixed as displayed in Fig. 12.10 (a)-(b). The quadrupole splitting progressively increases from 10 kHz at $V_{SG1}$ = -0.7 V to 25 kHz at $V_{SG1}$ = -1.1 V. Now to access region (iii), where we expect a constant strain field far underneath split metal gate 1, tuning $V_{CG}$ = -0.65 V and $V_{SG2}$ = -1.1 V are needed. There, the quadrupole splitting of about 10 kHz is detected and unchanged throughout the $V_{SG1}$ bias range of interest. Finally, when we set $V_{CG}$ = -0.2 V and $V_{SG2}$ = -2.52 V, we can access region (i) half-way in between center metal gate and split metal gate 1 where a maximum strain field is expected. The detected quadrupole splitting is about 45 kHz with all resonance peaks are clearly separated. Another lesson than can be drawn from this experiment by Fauzi et al [Fauzi2019] is that a slight change in the bias voltage condition can consideralby change the strain in the channel. We expect similar thing to happen for a device with more complex gate architecture.

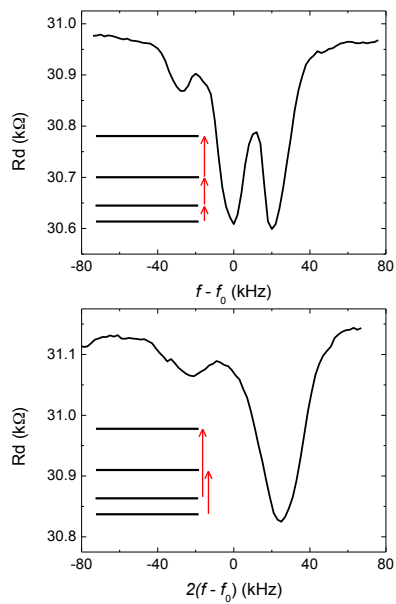

Fig. 12.11. $^{75}$As RDNMR spectra measured at a fundametal $f_0$ = 33.049 Hz (upper panel) and twice the fundamental frequency (lower panel). Inset in each panel shows its possible quantum transition. The spectra are measured at $B$ = 4.5 T and $T$ = 300 mK.

## 7. Overtone RDNMR

Overtone NMR, first proposed by Tycko and Opella, is a non-linear effect arising due to interaction between a quadrupole nuclei and rf magnetic field [Tycko1987]. The transition is observed at twice the Larmor frequency and is excited when the rf power is high enough so that a second-order perturbation starts kicking in. The selection rule $\Delta m = \pm 2$ normally forbidden in a normal setting is now relaxed in the overtone NMR due to the non-linearity effect. Overtone is categorically different in its origin from nuclear electric resonance, although both of them measure double quantum transition ($\Delta m = \pm 2$). Overtone is magnetic interaction while nuclear electric resonance is electrical interaction.

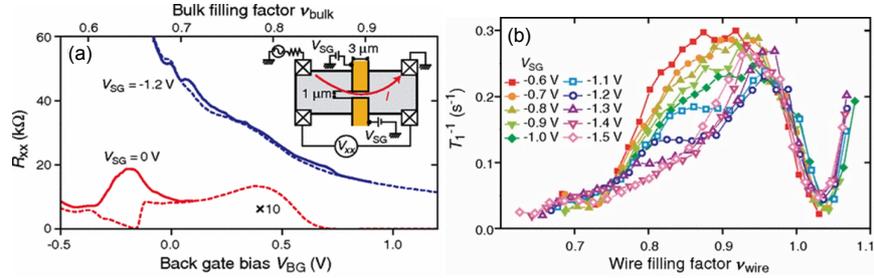

Fig. 12.12 (a) Longitudinal resistance Rxx as a function of back gate bias voltage with ($V_{SG}$ = -1.2 V) and without ($V_{SG}$ = 0 V) constriction formed. Fractional filling 2/3 is used to initialize and readout the nuclear spin polarization. (b) Nuclear spin relaxation rate, $T_1^{-1}$, as a function of calculated wire filling factor under various $V_{SG}$ bias voltages. The measurements are carried out at $B$ = 8.8 T and a base temperature $T$ = 50 mK. Reprinted with permission from [Kobayashi2011].

Fig. 12. 11 compares the fundamental and overtone RDNMR spectra with the rf power delivered at the top of cryostat of -30 dBm (1 μW). The central transition normally seen in the fundamental spectrum now disappears in the overtone RDNMR. Although the power is seemingly very low, the effective power at the sample could be much lower than that due to the impedance mismatch, surpisingly it still excites the overtone transition. Further investigation on power dependence is clearly needed to know where the overtone spectrum vanishes and whether the fundamental RDNMR spectrum would follow similar fate as well.

## 8. Spin dynamics in 1D semiconductor devices

Nuclear spin relaxation rate (1/T1) measures the rate at which nuclear spins dumps away their energy to the surrounding (e.g crsytal lattice and electrons) and equilbriate with. At low temperature where the lattice is nearly frozen, the relaxation is predominanly assisted by electrons. One then can learn a great deal of electron spin dynamics by looking at how fast the nuclear spin relaxes towards equilbrium.

**Skyrmions vs lateral confinement**

Shondi *et al* predicted in his seminal paper that spin excitation at filling factor $v$ = 1 quantum Hall ferromagnet involves not only a single spin reversal but collective spin reversal known as Skyrmions [Shondi1993].

The existence of Skyrmion spin excitation was later confirmed by Barret et al using optically pumped NMR technique [Barret1995]. They observe a sudden drop in the electron spin polarization as the filling factor is tuned slightly away from the exact filling factor $v$ = 1. At sufficiently low temperature, Skyrmion is also predicted to get crystalized whose energy is gapless at long wavelength limit. A strong coupling to the nuclei is then expected in this case. Indeed a rapid nuclear spin relaxation rate observed in a 2D system by a number of independent studies confirmed the prediction [Tycko1995, Smet2002, Hashimoto2002].

Kobayashi *et al* takes on the idea further to a quasi-one dimensional channel [Kobayashi2011]. The lateral confinement has a similar effect to raising the temperature, namely breaking the crystal long-range ordering. In a 2D system, there are many pathways to transmit information from one end to another end, but in a 1D system the pathways is restricted and the ordering is therefore prone to fluctuations. The Skyrmion crystal is then expected to melt by the confinement. If the assertion is true, the change from crystalline to liquid state should be reflected in the nuclear spin relaxation rate measurement.

To prove the hypothesis, they use a 3 $\mu$m-long split metal gates in a 2D electron gas embedded beneath the surface (inset of Fig. 12.14(a)). The device is also conveniently equipped with back gate to control global electron density. Similar to the previous report by Hashimoto *et al* [Hashimoto2002], they use fractional filling factor $v$ = 2/3 to initialize and readout resistively the remaining nuclear spin polarization due to interaction with Skyrmions confined in the channel as displayed in Fig. 12.14(a).

The nuclear spin relaxation rate as a function of calculated local filling factor measured at several bias voltage condition to the metal split gate is depicted in Fig. 12.14(b). The more negative bias applied to the split metal gate $V_{SG}$, the stronger the confinement becomes. For a less confined wire with $V_{SG}$ = -0.6 V, the relaxation rate is very fast at around $v_{wire}$ = 0.9 of about 0.3 s$^{-1}$, in agreement with previous reports in a 2D system. But then with increasing the confinement strength, the relaxation rate gets suppressed about 0.1 s$^{-1}$ at $V_{SG}$ = -1.5 V.

The study provides the first observation of the confinement effect on the nuclear spin relaxation rate due to interaction with the Skyrmion. However, given that there are no rigorous theoretical calculation to date, it is still not clear whether the suppression by a factor of 3 is expected at all when the Skyrmion is changed its phase from crystalline to liquid state.

**The 0.7 conductance anomaly**

The 0.7 conductance anomaly collectively refers to a zero-field conductance structure developed below the last integer plateau [Thomas1996, Cronenwett2002]. The observed features are believed due to a manifestation of electron-electron interactions in a quasi 1D channel. However, what kind of electronic state is emerged is still debated to date. The community is roughly divided into two distinct camps. First is a spin camp with its variance namely Kondo-related effect [Cronenwett2002, Iqbal2013] and spontaneous spin polarization [Thomas1996]. A second camp is a non-spin camp whose stories are revolved around a modified van-Hove singularity by Coulomb interaction [Bauer2013].

Cooper and Triphati [Cooper2008] gives a theoretical prediction that measuring the nuclear spin relaxation rate would give a promising test bed for discriminating a number of possible scenario leading to the 0.7 anomaly. However the prediction is still yet to be verified experimentally since generating non-equlibrium nuclear spin polarization close to zero-field in a quantum point contact is notoriously difficult to do. Recently, Fauzi *et al* proposes a method to generate non-equilibrium nuclear spin polarization that might be suited for the 0.7 anomaly study using a higher Landau level state [Fauzi2018].

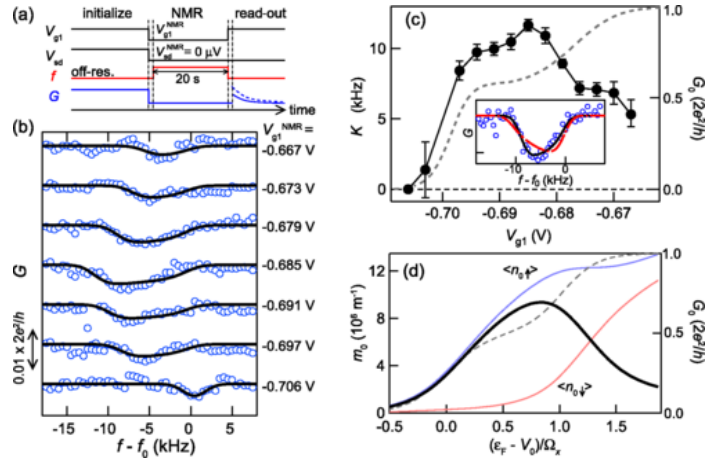

Fig. 12.13 (a) Schematic sequence for the pump-probe RDNMR measurement. (b) $^{75}$As RDNMR spectra collected at several bias voltages. (c) Knight shift $K$ plotted as a function of split gate bias below the first integer plateau. (d) Calculated magnetization density at the QPC center. Reprinted with permission from [Kawamura2015].

## 9. Electron spin polarization in a 1D system

We have discussed the reciprocality of the hyperfine interaction in section 4. The reciprocality causes a finite electron spin polarization $P$ to be picked up by the nuclear spin as an extra magnetic field known as Knight shift. For a 3/2 nuclear spin system, its energy level will be modified as schematically displayed in Fig. 12.8(b). Lets take a closer look at the central transition (1/2 → -1/2) so that we can exclude the influence from the first-order quadrupole interaction. The central resonance will be Knight shifted to a lower frequency by the amount given by

$$K_s = \frac{Au_0}{2h} \frac{n}{w_y w_z} P$$

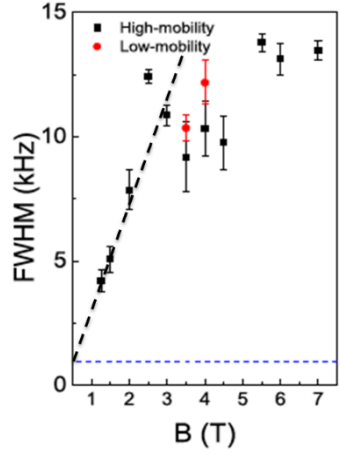

Fig. 12.14. $^{75}$As central transition resonance linewidth as a function of magnetic field. The black dashed line is the expected linear trend and the blue horizontal dashed line is the nuclear dipole limit of 1.5 kHz. Reprinted with permission from [Noorhidayati2020]

here the new notation $u_0$ is the semiconductor unit cell, $w$ is the effective confined region. The normalized electron spin polarization $P$ is proportional to the difference between the spin-up and spin-down electron density(transmittion) probability

$$P = \frac{n_\uparrow - n_\downarrow}{n_\uparrow + n_\downarrow} = \frac{T_\uparrow - T_\downarrow}{T_\uparrow + T_\downarrow}$$

Due to a direct proportionality, one can extract the electron spin polarization from the collected Knight shift RDNMR spectra.

In an attempt to solving the enigmatic 0.7 anomaly, Kawamura *et al* successfully applies RDNMR pump-probe technique to measure electron spin polarization in a quantum point contact under parallel magnetic field $B$ = 4.7 T and $T$ = 20 mK. They collects RDNMR spectra at several bias voltages where electrons only occupy the last 1D subband as displayed in Fig. 12.13(b) and then plot the extracted Knight shift as a function of bias voltage in Fig. 12.13(c). The data reveal that the Knight shift reaches a maximum value of around 10 kHz at the 0.7 regime and then drops down to zero when the conductance is pinched off ($G$ = 0) since there is no electrons to participate. The Knight shift drops as well when the conductance hits the integer plateau ($G = 2e^2/h$). Converting the Knight shift to the polarization, they are able to estimate the polarization reaching 70% at the 0.7 regime. Moreover, the collected Knight shift profile is continously evolved in agreement with numerical simulation without bound state formation depicted in Fig. 12.13(d).

Note that the Knight shifts itself are collected in a condition where the spin degeneracy is almost fully lifted by the action of magnetic field of $B = 4.7$ T. It remains to be seen whether the conclusion of no bound state formation holds up still at zero magnetic field or close to.

Similar RDNMR pump-probe technique has been used to investigate breakdown mechanism of electrons at a quantum point contact filling factor $v = 1$ quantum Hall effect [Chida2012, Hashisaka2020]. The electrons are subjected to various bias conditions from weak to strong bias regime. The collected Knight shift monotonically decreases with increasing bias voltage. This is reasonable since the inter-edge spin flip scattering (forward or backward) probability increases as well, making the effective electron spin polarization reduces. Interestingly, they found that the electron spin polarization exctracted from the Knigh shift and shot noises differs at high bias regime, namely the shot noise is saturated. They suggest that the discrepancy is due to additional breakdown mechanism namely the closure of exchange-enhanced spin gap.

In addition to measuring the Knight shift, one can also measure the central transition linewidth. This would give a new piece of information on how electrons distribute themselves in the point contact when subjected to a perpendicular magnetic field.

Fig. 12.14 displays the central transition linewidth measured at 1.25 up to 7 T. Naively speaking, one would expect the central transition linewidth to scale linearly with the field indicated by the black dashed line in Fig. 12.14. This is because to maintain the bulk filling factor v = 2 at different magnetic field, one has to tune the electron density $n$. While this is true for relatively low to moderate field below 3 T, it deviates substantially from a linear fit when the field is ramped above.

When the field is ramped up, the Coulomb interaction between particles increases in proportion to the square root of magnetic field strength. To minimize the Coulomb interaction, the particles have to distance themselves from each other.

## Acknowledgments

We would like to thank Katsushi Hashimoto, Bhaskaran Muralidharan, and Tomosuke Aono for fruitful discussions on the subject and collaborations. We would like to thank (late) Katsume Nagase and Ken Sato for their assistance in fabrication and technical processes. We would like to thank a number of Master and PhD students directly involved in the project throughout the years M. F. Sahdan, S. Maeda, M. Takahashi, A. Noorhidayati, and T. Sobue.

## References

[Lee01] S. C. Lee, K. Kim, J. Kim, S. Lee, Yi J. Han, S. W. Kim, K. S. Ha, C. Cheong, J. Magn. Reson. **150**, 207 (2001).

[Ono04] Keiji Ono and Seigo Tarucha, Phys. Rev. Lett. 92, 256803 (2004).

[Chekhovich13] EA Chekhovich, MN Makhonin, AI Tartakovskii, Amir Yacoby, H Bluhm, KC Nowack, LMK Vandersypen, Nature Materials **12**, 494-504 (2013).

[Rugar04] D. Rugar, R. Budakian, H. J. Mamin, B. W. Chui, Nature **430**, 329 (2004).

[Dobers1988] M. Dobers, K. v. Klitzing, J. Schneider, G. Weimann, and K. Ploog, Phys. Rev. Lett. **61,** 1650 (1988).

[Li2008] Y. Q. Li and J. H. Smet, Nuclear-Electron Spin Interactions in the Quantum Hall Regime, in Spin Physics in Semiconductors, edited by Michel I. Dyakonov (Springer, Berlin/Heidelberg, 2008), pp. 347-388.

[Gervais2009] G. Gervais, Resistively detected NMR in GaAs/AlGaAs, in Electron Spin Resonance and Related Phenomena in Low-Dimensional

Structures, edited by M. Fanciulli (Springer,Berlin/Heidelberg, 2009), pp. 35–50.

[Hirayama2009] Y. Hirayama, G. Yusa, K. Hashimoto, N. Kumada, T. Ota, andK. Muraki, Semicond. Sci. Technol. **24**, 023001 (2009).

[Duprez2019] H. Duprez, E. Sivre, A. Anthore, A. Aassime, A. Cavanna, A. Ouerghi, U. Gennser, and F. Pierre, Phys. Rev. X **9**, 021030 (2019).

[vanWees1988] B. J. van Wees, H. van Houten, C. W. J. Beenakker, J. G. Williamson, L. P. Kouwenhoven, D. van der Marel, and C. T. Foxon, Phys. Rev. Lett. **60**, 848 (1988).

[Wharam1988] D. A. Wharam, T. J. Thornton, R. Newbury, M. Pepper, H. Ahmed, J. E. F. Frost, D. G. Hasko, D. C. Peacock, D. A. Ritchie, and G. A. C. Jones, J. Phys. C: Solid State Phys. **21** L209 (1988).

[Thornton1986] T. J. Thornton, M. Pepper, H. Ahmed, D. Andrews, and G. J. Davies, Phys. Rev. Lett. **56**, 1198 (1986).

[Buttiker1990] M. Buttiker, Phys. Rev. B **41**, 7906 (1990).

[Fauzi2018] M. H. Fauzi, A. Noorhidayati, M. F. Sahdan, K. Sato, K. Nagase, and Y. Hirayama, Phys. Rev. B **97**, 201412(R), (2018).

[Paget1977] D. Paget, G. Lampel, B. Sapoval, and V. I. Safarov, Phys. Rev. B **15**, 5780 (1977).

[Dixon1997] David C. Dixon, Keith R. Wald, Paul L. McEuen, and M. R. Melloch, Phys. Rev. B **56**, 4743 (1997).

[Haug1993] R J Haug, Semicond. Sci. and Technol **8**, 131 (1993).

[Fauzi2017] M. H. Fauzi, A. Singha, M. F. Sahdan, M. Takahashi, K. Sato, K. Nagase, B. Muralidharan and Y. Hirayama, Phys. Rev. B **95**, 241404(R), (2017).

[Wald1994] Keith R. Wald, Leo P. Kouwenhoven, Paul L. McEuen, Nijs C. van der Vaart, and C. T. Foxon, Phys. Rev. Lett. **73**, 1011 (1994).

[Kane1992] B. E. Kane, L. N. Pfeiffer, and K. W. West, Phys. Rev. B **46**, 7264(R) (1992).

[Singha2017] Aniket Singha, M. H. Fauzi, Yoshiro Hirayama, Bhaskaran Muralidharan. Phys. Rev. B **95**, 115416, (2017).

[Stano2018] Peter Stano, Tomosuke Aono, and Minoru Kawamura, Phys. Rev. B **97**, 075440 (2018).

[Yusa2005] G. Yusa, K. Muraki, K. Takashina, K. Hashimoto and Y. Hirayama, Nature, **434**, 1001 (2005).

[Ota2007] T. Ota, G. Yusa, N. Kumada, S. Miyashita, and Y. Hirayama, Appl. Phys. Lett. **90**, 102118 (2007).

[Corcoles2009] A. Córcoles, C. J. B. Ford, M. Pepper, G. A. C. Jones, H. E. Beere, and D. A. Ritchie, Phys. Rev. B **80**, 115326 (2009).

[Miyamoto2016] S. Miyamoto, T. Miura, S. Watanabe, K. Nagase, Y. Hirayama, Nano Lett. **16**, 1596-1601 (2016).

[Noorhidayati2020] A. Noorhidayati, M. H. Fauzi, M. F. Sahdan, S. Maeda, K. Sato, K. Nagase, and Y. Hirayama. Phys. Rev. B, **101**, 035425 (2020).

[Desrat2002] W. Desrat, D. K. Maude, M. Potemski, J. C. Portal,Z. R. Wasilewski, and G. Hill, Phys. Rev. Lett. **88**, 256807 (2002).

[Sundfors1969] R. K. Sundfors, Phys. Rev. **177**, 1221 (1969).

[Taylor1959] E. F. Taylor and N. Bloembergen, Phys. Rev. 113, 431 (1959).

[Fauzi2019] M. H. Fauzi, M. F. Sahdan, M. Takahashi, A. Basak, K. Sato, K. Nagase, B. Muralidharan, and Y. Hirayama, Phys. Rev. B **100**, 241301(R) (2019).

[Tycko1987] R. Tycko and S. J. Opella, J. Chem. Phys. **86**, 1761 (1987).

[Kobayashi2011] Takashi Kobayashi, Norio Kumada, Takeshi Ota, Satoshi Sasaki, and Yoshiro Hirayama, Phys. Rev. Lett. **107**, 126807 (2011).

[Shondi1993] S. L. Sondhi, A. Karlhede, S. A. Kivelson, and E. H. Rezayi, Phys. Rev. B **47**, 16419 (1993).

[Barret1995] S. E. Barrett, G. Dabbagh, L. N. Pfeiffer, K. W. West, and R. Tycko, Phys. Rev. Lett. **74**, 5112 (1995).

[Tycko1995] R Tycko, SE Barrett, G Dabbagh, LN Pfeiffer, KW West, Science **268**, 1460-1463 (1995).

[Hashimoto2002] K. Hashimoto, K. Muraki, T. Saku, and Y. Hirayama, Phys. Rev. Lett. **88**, 176601 (2002).

[Smet2002] J. H. Smet, R. A. Deutschmann, F. Ertl, W. Wegscheider, G. Abstreiter, and K. von Klitzing, Nature **415**, 281–286 (2002).

[Thomas1996] K. J. Thomas, J. T. Nicholls, M. Y. Simmons, M. Pepper, D. R. Mace, and D. A. Ritchie, Phys. Rev. Lett. **77**, 135 (1996).

[Cronenwett2002] S. M. Cronenwett, H. J. Lynch, D. Goldhaber-Gordon, L. P. Kouwenhoven, C. M. Marcus, K. Hirose, N. S. Wingreen, and V. Umansky, Phys. Rev. Lett. **88**, 226805 (2002).

[Iqbal2013] M. J. Iqbal, Roi Levy, E. J. Koop, J. B. Dekker, J. P. de Jong, J. H. M. van der Velde, D. Reuter, A. D. Wieck, Ramón Aguado, Yigal Meir & C. H. van der Wal, Nature **501**, 79 (2013).

[Bauer2013] Florian Bauer, Jan Heyder, Enrico Schubert, David Borowsky, Daniela Taubert, Benedikt Bruognolo, Dieter Schuh, Werner Wegscheider, Jan von Delft & Stefan Ludwig, Nature **501**, 73 (2013).

[Cooper2008] N. R. Cooper and V. Tripathi, Phys. Rev. B **77**, 245324 (2008).

[Kawamura2015] Minoru Kawamura, Keiji Ono, Peter Stano, Kimitoshi Kono, and Tomosuke Aono, Phys. Rev. Lett. **115**, 036601 (2015).

[Chida2012] Kensaku Chida, Masayuki Hashisaka, Yoshiaki Yamauchi, Shuji Nakamura, Tomonori Arakawa, Tomoki Machida, Kensuke Kobayashi, and Teruo Ono, Phys. Rev. B **85**, 041309(R) (2012).

[Hashisaka2020] Masayuki Hashisaka, Koji Muraki, and Toshimasa Fujisawa, Phys. Rev. B **101**, 041303(R) (2020).